# Georeferenced lives

A.C. Sparavigna[1] and R. Marazzato[2]
1 Department of Applied Science and Technology, Politecnico di Torino, Italy
2 Department of Control and Computer Engineering, Politecnico di Torino, Italy

*To give a georeference means to give a reference as existing in the physical space of Earth. This procedure is widely used for the location of archaeological, historical and other sites when geographic information systems (GIS) are used. Here we are proposing to georeference the lives of famous people (in the paper, Newton and Schiaparelli) for teaching purposes, to increase the appeal of some scientific disciplines.*

Researches on the georeferencing methodologies for the localization and evaluation of cultural heritage are well known and in strong increasingly development [1]. To georeference something, that is to give its georeference, means to define its existence in the physical space, in particular that of the Earth. On a finer scale, georeferencing is used to have a precise mapping of the structures in an archaeological or historical site [2,3]. This procedure is imperative for the data modelling of geographic information systems (GIS). In a wider sense, the research on georeferencing can be considered as concerning the location of some sites relevant for cultural or even social activities on public maps, such as those of the web services (Google Earth, Wikimapia), as we have discussed in Ref. [4]. Besides of increasing the knowledge of our past, georeference contributes can be used for studying and improving some sustainable developments and maintenance policy of archaeological sites. In addition, this can be an economical resource for local population. What we are suggesting in this short paper is to attach time to a georeference, or, in the same manner, to attach georeferences to a timeline, with the aim of increasing the appeal of some scientific disciplines.

In fact, a georeference can be quite interesting for teaching purposes. The use of Google Earth for teaching, that is, a use of a georeferencing approach to scientific disciplines, was already reported in several papers [5-9]. One of these references is quite interesting for our actual discussion, because it is proposing to stimulate the students to study physics, by bringing to life the stories of famous scientists by paying a "flying visit to their homes". For instance, the author suggests to visit Woolsthorpe Manor, the birthplace of Sir Isaac Newton.

If we were teaching history, we could decide, for instance, to create a web site concerning life and military campaigns of Julius Caesar. The site can be accompanied by a browser interface of Google Earth allowing users to navigate the Earth's surface. The chronology of Caesar's life is in such a manner translated in all the places where he lived and visited. The places can be recorded and then shared, by saving them into "placemarks" files. In the case that we use the free version of Google Earth, the data can be saved in the KML format. Keyhole Markup Language is the notation of a mark-up language (XML) for expressing geographic annotation and visualization within Internet-based, two-dimensional maps and three-dimensional Earth browsers [4].

Of course, implementing Caesar's life is a huge task. For sure, it would be great to have the Gallic Wars [9], visualized on Google Earth with their proper timelines. Here we are bringing two examples, not as complex as Julius Caesar's life, that are about the life of Sir Isaac Newton and of the Egyptologist Ernesto Schiaparelli. That is, here we have as timelines the lives of two great scientists, georeferenced on maps.

**The places of Sir Isaac Newton's life**
Since our aim is the proposal of a method, not yet the implementation of a real site let us use one among the simplest maps and satellite services that we can find on the web. This is the ACME Mapper [10]. It is based on the same satellite imagery of Google Earth. ACME Mapper has the possibility to mark the places, and provides a list of all the marked places with their coordinates and distances. In Figure 1 for instance, the white markers are showing the places where Isaac Newton lived.

There are many facts on the Newton's life that most people do not know. However, we have Wikipedia helping us in any preliminary approach to the study of his life [11]. On the map of Fig.1, we see, near Nottingham, three markers showing where he was born and studied and where he came back during the Great Plague. It is there, in a garden near Nottingham, that he had the intuition of a universal gravitation observing a falling apple [12]. The other markers are indicating Cambridge, London and Southampton. In London, one of the markers is located at the Tower of London, because during the last twenty years of his life, Newton was Master of the Mint, and the Mint was there at that time.

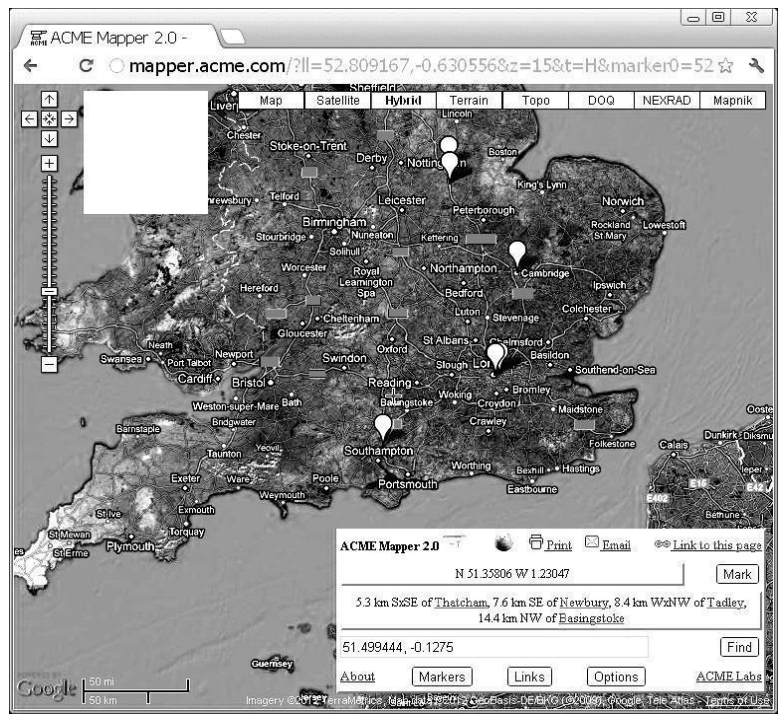

Fig.1. The white markers indicate where Sir Isaac Newton lived

Actually, the map in Fig.1 is representing a possible interactive map, where zooming on the places is allowed, and may be, different labels or a color code corresponding to a timeline can be used. It turns to be a good mnemonic map, to illustrate Newton's life, based on an approach different from other maps [13].

**Ernesto Schiaparelli's sites, georeferenced.**
Again, let us use Wikipedia for the life of Ernesto Schiaparelli (1856-1928) [14].
He was born in Occhieppo Inferiore (Biella, Italy). He was the archaeologist who found the Queen Nefertari's tomb at Deir el-Medina, in the Valley of the Queens, in 1904. He excavated the tomb of the royal architect Kha and his wife Merit, in 1906 [15,16]. This tomb was found intact and all the funerary items are in Turin, at the Egyptian Museum.

Schiaparelli was the director of this Museum, after his activity as director of the Egyptian Museum in Florence. For Egyptology, the Egyptian Museum of Turin is considered as the second museum in the world, after that of Cairo, because of the rich collections it possesses, strongly increased by the Schaparelli's several seasons of excavations in Egypt. Senator of the Kingdom of Italy, he founded the Association to Succour Italian Missionaries in Luxor, with the aim of relieving their poverty.

From 1903 to 1920, Schiaparelli undertook twelve archaeological campaigns, opening sites in Giza, Hermopolis, Assiut, Qau el-Kebir, Gebelien and Aswan. In Figure 2 we see a detail of the georeferences to Schiaparelli's life, those in Egypt. Of course, there are the locations in Italy too.

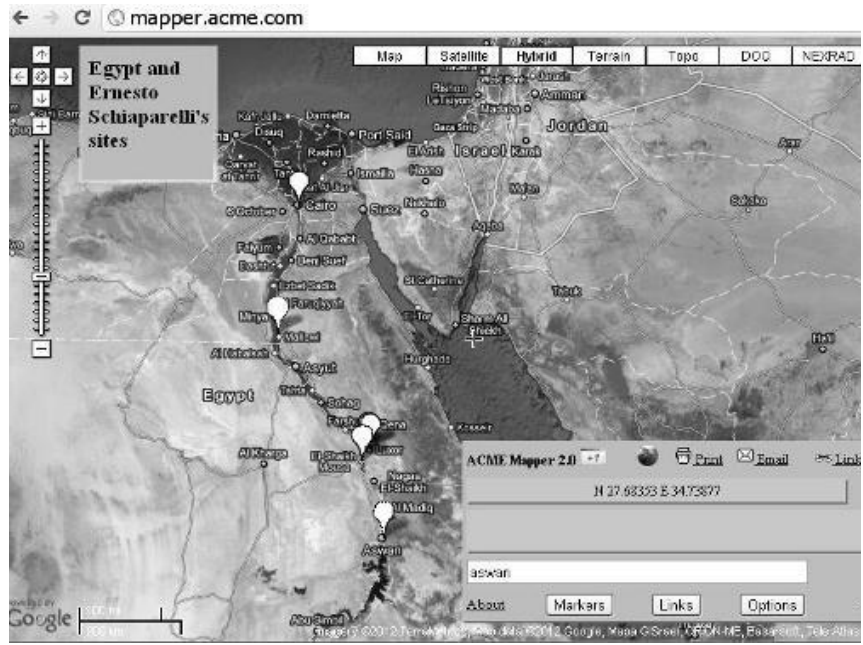

Fig.2. Ernesto Schiaparelli's sites in Egypt marked by means of ACME Mapper.

**Discussion**

Figures 1 and 2 are the georeferences on a map where these persons lived and worked. The maps can be improved using icons or labels to give the timelines. In these two examples, which are just a simplified illustration of the method, we used an approach with the ACME Mapper: a more complex XML application is under development to shows the Schiaparelli's tours in Egypt.

It could be questionable the necessity of such tools, but, our opinion is that it is quite helpful for teaching and mnemonic purposes. As told in Ref.3, any application, which is able to give georeferences, can help and improve the study and comparative analysis of our cultural heritage. Moreover, the same reference remarks that the availability of some free tools for manipulating and processing images is the key that will intensify the use of georeferencing for historical, or more generally, cultural maps.

Georeferences are quite familiar to researches using the GIS systems. As discussed in Ref.4, a popular example of GIS is Google Earth, which is offering the possibility of virtual tours to the users of the Web. The service provides, besides the labels of usual maps, the icons of Wikipedia. Moreover, dragging the icon of Street View on the map, several dots and lines appear in blue on it, each dot corresponds to a picture uploaded by the users and the line to a panoramic view. The blue dots are an example of georeferencing pictures (Fig.3). Using these images, pictures and movies, we can check the landscape how it is, and compare with its aerial view. It is then quite clear that a GIS service is fundamental to have a precise location of the features of a cultural landscape.

Of course, images are not confined to be some pictures of a landscape. In the case of a site devoted to the study of the Schiaparelli's excavations in Egypt, we could for instance add the map with some pictures of the Nefertati's tomb or of the Kha's tomb, and of items found inside it. They could be images or documents in PDF files too. The appearance of such georeferencing could be as shown in Fig.3: after dragging Street View icon on the Deir el-Medina location, several dots appear.

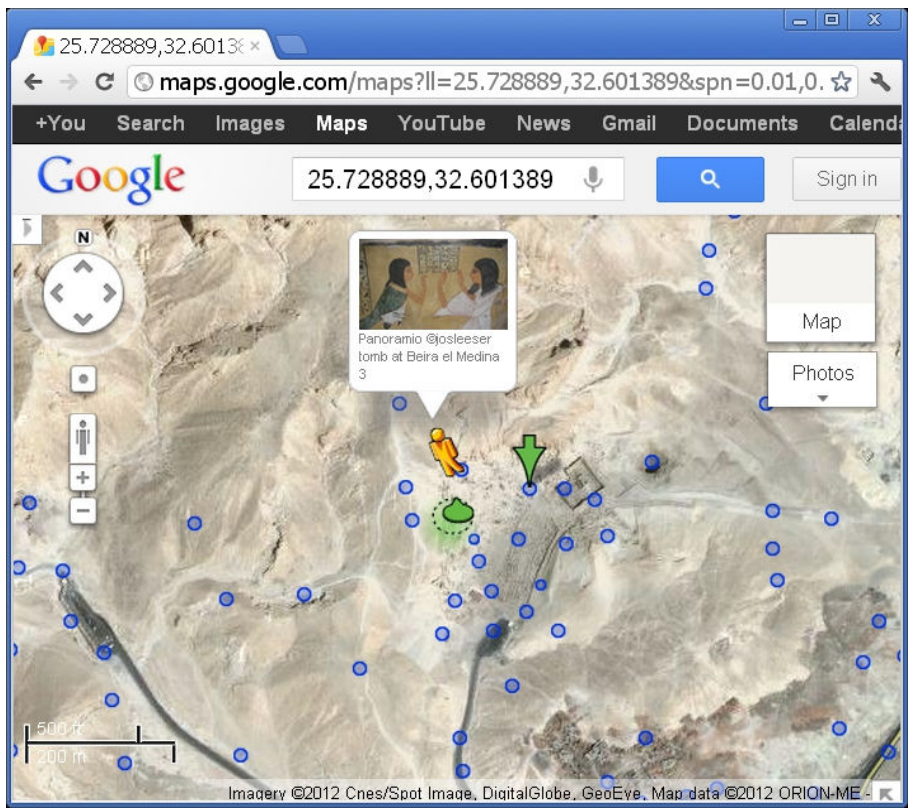

Fig.3. An example of georeferencing pictures of Deir el-Medina site.

The ancient roman had some *itineraria,* which were maps in the form of a listing of cities, villages and even refreshment places, with intervening distances. By means of them, the ancient travellers had some idea of how they had to move, to get where they had to go. The satellite maps, as we have proposed, can become our *itineraria* in the life of great people.